\documentclass[aps,superscriptaddress,nofootinbib,eqsecnum,prd,notitlepage,twocolumn,showkeys]{revtex4-1} 

\pdfoutput=1

\usepackage{amsfonts}
\usepackage{amsmath}
\usepackage{amssymb}
\usepackage{graphicx,color}
\usepackage{float}
\usepackage{hyperref}
\usepackage{subfigure}
\usepackage{dcolumn}
\usepackage{soul}
\usepackage{ulem}
\usepackage{verbatim}
\newcommand{\beq}{\begin{equation}\begin{aligned}}
\newcommand{\eeq}{\end{aligned}\end{equation}}

\begin{document}

\title{Dissipative Dark Energy can explain the DESI phantom crossing }
\author{Prolay Chanda}
\email{prolay.chanda@tifr.res.in }
\affiliation{Tata Institute of Fundamental Research, Homi Bhabha Road, Mumbai 400005, India}

\author{Subinoy Das}
\email{subinoy@iiap.res.in}
\affiliation{Indian Institute of Astrophysics, Bengaluru, Karnataka 560034, India}

\author{Suratna Das}
\email{suratna.das@ashoka.edu.in}
\affiliation{Department of Physics, Ashoka University,
   Rajiv Gandhi Education City, Rai, Sonipat 131029, Haryana, India}

\begin{abstract}

DESI results preferring an evolving dark energy component that appears to cross the phantom-divide in the recent past has raised a lot of interest in exploring the nature of dark energy. We present here a simple dissipative dark energy scenario that can explain both the evolving nature of dark energy as well as its crossing of the phantom-divide without invoking any pathological phantom-like dynamics for the quintessence field. We show that even weak dissipation of the quintessence is enough to explain the current DESI observations. 

\end{abstract}

\maketitle

\section{Introduction}

The concurrent standard model of our Universe, a.k.a. the $\Lambda$CDM model, is currently encountering tensions with the latest high-precision cosmological data, among which the persistent $H_0$ tension \cite{Riess:2016jrr, Riess:2021jrx} is worth mentioning. The $\Lambda$CDM model is named after the two dominant cosmological fluids of the present cosmic plasma, namely the Cosmological Constant ($\Lambda$) that explains the current accelerating expansion of our Universe, and the Cold Dark Matter (CDM) that gives rise to the structures we see in our Universe. Although the Cosmological Constant (CC) as a Dark Energy (DE) candidate poses the most extreme fine-tuning problem in physics where the mismatch between the observational value and the theoretically value predicted from quantum field theory is of 120 orders in magnitude \cite{Zeldovich:1968ehl, Weinberg:1988cp}, the cosmological observations up to the Planck data release in 2018 had largely remained consistent with a CC having a {\it constant} equation of state (EoS) $w_0\sim\,-1$ \cite{Planck:2018vyg}. Yet, due to the elusive origin and nature of DE, it was motivated to parametrize the EoS of the DE that deviates from the constant EoS picture of CC. The most explored parameterization of this type is the CPL parametrization \cite{Chevallier:2000qy, Linder:2002et} where the EoS is made a function of the scale factor $a$ as $w(a)=w_0+w_a(1-a)$ with $w_0$ and $w_a$ being two constants, implying that at $a=1$, i.e., today, $w(a)=w_0\sim-1$.\footnote{Several other parameterizations of DE EoS have also been explored in the literature, e.g., see \cite{Corasaniti:2002vg, Sharma:2022ifr}.} It is only very recently that the baryon acoustic oscillation (BAO) measurements by the Dark Energy Spectroscopic Instrument (DESI) (both DR1 \cite{DESI:2024uvr, DESI:2024lzq} and DR2 \cite{DESI:2025zgx}) have started hinting towards an evolving equation of state $w(z)$ (here $z$ represents the redshift, and $1+z=1/a$) for DE at a 2-$\sigma$ confidence level, with the CC lying outside the 2-$\sigma$ contour.  DESI data in combination with several SN Ia and CMB data infer that $w_0>-1$ and $w_a<0$. This result indicates that today $w_0>-1$, and $w(z)$ decreases in the past with increasing $z$ crossing $w=-1$ in some recent past. If these results stand the test of time, then they will clearly indicate  a dynamical DE scenario rather than it being just a mere constant (i.e., the CC). Other independent analysis also support the dynamical DE picture, e,g, see \cite{Cline:2025sbt, deCruzPerez:2025dni}. 

In the simplest scenario, a slowly rolling scalar field $\phi$ with canonical kinetic term, dubbed quintessence, yields dynamical DE \cite{Ratra:1987rm, Tsujikawa:2013fta} in which case the EoS of the quintessence field remains $w_\phi\geq-1$. However, the DESI results also suggest  \cite{Caldwell:1999ew,Tada:2024znt} that the quintessence had been {\it phantom}-like in the past before it had crossed the {\it phantom-divide} ($w=-1$) in the recent past. Such an outcome is perplexing as theoretically phantom dynamics is pathological as it calls for non-canonical physics such as negative kinetic energy \cite{Caldwell:2003, Ludwick:2017tox} that makes the Hamiltonian unbounded from below. Though the standard thawing quintessence models within GR cannot explain such {\it phantom-crossing} \cite{Wolf:2024eph}, it was shown in \cite{Wolf:2025jed} that non-minimally coupled quintessence thawing models are favored by the data. However, such non-minimally coupled models yield fifth forces on scales smaller than the cosmological scales, which have not been observed yet \cite{Wolf:2025jed}. Ghost-free quintessence thawing models on a braneworld (propagating on (4+1) dimensions) have been shown to be able to tackle the observed phantom-crossing more amicably in \cite{Mishra:2025goj}. Several modified gravity theories have also been explored in the same context \cite{Ye:2024ywg, Odintsov:2024woi, Pan:2025psn, Yao:2025wlx, Yang:2025mws, Nojiri:2025low, Tsujikawa:2025wca, Efstratiou:2025iqi}. Other plausible scenarios, e.g., new physics prior to recombination \cite{Gonzalez-Fuentes:2026rgu}, negative quintessence \cite{Gomez-Valent:2025mfl, Gonzalez-Fuentes:2025lei} and coupling of the quintessence with vector fields \cite{Tsujikawa:2026xqm}, have also been explored to address the DESI observed phantom-crossing. 

A simpler scenario that explains the observed phantom-crossing without invoking any phantom dynamics is the interacting DM-DE scenario \cite{Amendola:1999er, Amendola:2003eq}, where the EoS inferred from cosmological observations is an effective quantity, $w_{\rm eff}$, rather than the intrinsic quintessence EoS, $w_\phi$, as shown in \cite{Das:2005yj}. The reason is that an observer unaware of the DM-DE interactions typically assumes the standard CDM evolution (where its energy density dilutes with inverse-cubic power of the scale factor). However, the DM mass, depending on the quintessence field due to DM-DE interactions, yields a different scaling of the DM energy density. Thus, interpreting the resulting expansion history within a non-interacting framework therefore leads to an effective DE description characterized by $w_{\rm eff}$ that includes the impact of the modified DM evolution due to DM-DE interactions. It was shown in \cite{Das:2005yj} that this effective EoS, $w_{\rm eff}$, appears to have crossed the phantom-divide for an observer oblivious of the interactions between DM and DE. After the publication of the DESI results, such models were explored with renewed interest \cite{Chakraborty:2024xas, Chakraborty:2025syu, deCruzPerez:2025dni, Gomez-Valent:2026ept, Antusch:2026ldp, Chen:2025ywv, Wang:2026wrk}, and the interacting dark sector models were shown to be more robust than the non-interacting scenarios given the current precision data \cite{Li:2026xaz, Gomez-Valent:2026ept, Wang:2026kbg, Scherer:2025esj,Silva:2025hxw}. 

In this {\it letter}, we investigate a dissipative DE scenario and demonstrate that it can naturally account for the recent DESI observations. The slow-roll dynamics of the quintessence field that explains the late-time acceleration of our Universe is very similar to the slow-roll dynamics of the inflaton field that gives rise to the early-time accelerating epoch, a.k.a. cosmic inflation. Warm Inflation (WI) \cite{Berera:1995ie} is a variant inflationary scenario where the inflaton field keeps dissipating its energy to a coexisting subdominant radiation bath during the inflationary period. This is in contrast to the standard inflationary scenarios where the couplings of the inflaton field to other degrees of freedom are neglected or considered to be suppressed during inflation, and thus it is assumed that the inflaton will slow-roll in isolation to drive the accelerated expansion. For this very reason, in standard scenarios, the Universe needs to be ``reheated" post inflation \cite{Baumann:2009ds}. The advantage of maintaining a subdominant thermal bath during WI through dissipation is that when WI ends, the Universe smoothly transitions into a radiation-dominated epoch without any post-inflationary reheating epoch, the physics of which is still largely unknown and cannot be probed in observations. For recent reviews on WI, see \cite{Kamali:2023lzq, Berera:2023liv}.\footnote{It is to note that dissipative dynamics of the inflaton field have been explored in standard inflationary scenarios as well, where the particles created due to dissipation do not thermalize to yield a subdominant thermal bath as in WI \cite{Creminelli:2023aly}.} Such a feature of WI turns out to be an added advantage for quintessential inflation models in which the same scalar field plays the role of inflaton (in early times) and quintessence (at late times). This is due to the fact that in standard quintessential scenarios, as the inflaton has to play the role of quintessence at late times, it cannot decay away during the reheating epoch, and one then has to rely on non-standard reheating mechanisms, like gravitational reheating \cite{Ford:1986sy, Chun:2009yu}, instant preheating \cite{Felder:1998vq, Campos:2002yk}, curvaton reheating \cite{Feng:2002nb, BuenoSanchez:2007jxm}, non-minimal \cite{Dimopoulos:2018wfg} or Ricci reheating \cite{Bettoni:2018utf, Opferkuch:2019zbd}, etc. However, as WI doesn't call for a reheating epoch, the remnant inflaton field can naturally become quintessence at late times if one considers warm quintessential inflation models. Early studies of warm quintessential inflation \cite{Dimopoulos:2019gpz, Rosa:2019jci} considered scenarios in which the inflaton undergoes dissipative evolution during inflation, whereas the quintessence field retains standard non-dissipative dynamics. Later in \cite{Lima:2019yyv}, a model was constructed where the inflaton not only dissipates to a radiation bath but also to a matter bath during inflation, and though the dissipation to radiation gradually diminishes post-inflation, the dissipation to matter remains active till late times when the inflaton becomes quintessence. Such dissipation of matter fluid was shown to yield the right abundances of dark matter. Thus, this model unifies not only inflation and dark energy but also dark matter, i.e., all three elusive cosmological fluids. This is also the first instance where the concept of dissipative DE was coined. The dynamical stability of such a model was studied and confirmed in \cite{Das:2023rat}.\footnote{Another scenario of dissipative DE was explored in \cite{Berghaus:2020ekh, Berghaus:2023ypi} where DE dissipates to a concurrent Dark Radiation bath. Such models cannot explain the phantom-crossing of the DE EoS as in these models DM evolves as standard matter fluid.} Recently, it was also shown that when the dissipation is strong enough, the quintessence slow-roll dynamics can also be made to sustain along steep potentials \cite{Das:2025teu}. 

We begin in Sec.~\ref{sec:diss_DE} with a brief overview of the dissipative DE scenario. We then present our main results in Sec.~\ref{sec:results}, showing that this framework is consistent with the recent DESI observations by naturally yielding an evolving effective DE EoS  that explains the observed phantom-divide crossing. Finally, in Sec.~\ref{sec:summary}, we summarize the key features of the scenario, discuss the implications of our findings, and conclude.


\section{Model and Dynamics of dissipative Dark Energy}
\label{sec:diss_DE}

In the dissipative DE model, the quintessence field decays to a concurrent subdominant matter fluid that accounts for today's dark matter (DM). Following the dissipative dynamics of WI, we can write the evolution equations for the energy densities of the dissipative DE ($\rho_\phi$) and the concurrent DM fluid ($\rho_{\rm dm}$) as 
\begin{eqnarray}
&&\dot\rho_\phi+3H(\rho_\phi+p_\phi)=-\Upsilon_m\dot\phi^2, \label{quintessence-dyn} \\
&&\dot\rho_{\rm dm}+3H(\rho_{\rm dm}+p_{\rm dm})=\Upsilon_m\dot\phi^2. \label{dm-dyn}
\end{eqnarray}
It is evident from the above equations that in this dissipative DE scenario, the evolutions of DE and DM are intertwined (through the dissipative coefficient $\Upsilon_m$), which is an outcome of the conservation of the total stress-energy tensor. The energy density $\rho_\phi$ and the pressure $p_\phi$ of the quintessence field are the same as in the standard case: $\rho_\phi=\dot\phi^2/2+V(\phi)$ and $p_\phi=\dot\phi^2/2-V(\phi)$, and the DM is also considered as a pressure-less dust, i.e., $p_{\rm dm}=0$. The EoS of the quintessence field is defined in the standard way, too, as
\begin{eqnarray}
w_\phi=\frac{p_\phi}{\rho_\phi}.\label{w-phi}
\end{eqnarray}

Introduction of dissipation modifies the dynamics of the quintessence as well as that of the DM fluid. In the quintessence dynamics, the dissipation brings in added friction. This can be seen if one rewrites Eq.~(\ref{quintessence-dyn}) in terms of the equation of motion (EoM) of the quintessence field as 
\begin{eqnarray}
\ddot\phi+(3H+\Upsilon_m)\dot\phi+V,_\phi=0. \label{KG-eqn}
\end{eqnarray}
Here, the dissipation behaves as an extra friction term apart from the Hubble friction that is already present in the EoM due to the background expansion.\footnote{To compare the dissipative DE model with the interactive DE picture studied in \cite{Das:2005yj, Chakraborty:2024xas, Chakraborty:2025syu}, one can note that the Yukawa-type interactions introduced between quintessence and DM fluid modify the quintessence potential. The interactions don't bring in extra friction in interactive DE models like in the case of dissipative DE.} WI, having a similar dissipative inflaton dynamics, are often studied in two limiting cases -- one is the weak dissipative regime, where the Hubble friction is greater than the friction due to dissipation, and the other is the strong dissipative regime, where the dissipation friction governs the inflaton dynamics. In a similar way, the quintessence dynamics studied in \cite{Lima:2019yyv} takes place in a weak dissipative regime, whereas the one studied in \cite{Das:2025teu} is an example of strong dissipative dynamics. Defining a dimensionless quantity $Q_m$ as 
\begin{eqnarray}
Q_m=\frac{\Upsilon_m}{3H}, \label{Qm}
\end{eqnarray}
one can define weak (strong) dissipative dynamics when $Q_m\ll1$ ($Q_m\gg 1$). 

For the DM fluid, one can see from Eq.~(\ref{dm-dyn}) that energy is being transferred to it from quintessence through dissipation. As a result, in this dissipative DE scenario, the energy density $\rho_{\rm dm}$ of the DM fluid no longer scales as $1/a^3$ as it happens in the non-dissipative case.  The scaling of the DM fluid will depend on the form of the dissipative coefficient $\Upsilon_m$. The dissipation coefficient proposed in Ref.~\cite{Lima:2019yyv} that is relevant during the quintessence evolution takes the form
\begin{eqnarray}
\Upsilon_m=\frac{M_{\rm diss}^2}{\rho_{\rm dm}^{1/4}},
\label{ups-m}
\end{eqnarray} where $M_{\rm diss}$ is a constant with mass dimension one that sets the strength of the dissipation. In viable particle physics models of WI, the dissipative coefficient, $\Upsilon_r$, that dissipates energy from the inflaton fluid to the thermalized radiation bath, turns out to be a function of the inflaton field $\phi$ and the temperature of the radiation bath $T$. As the radiation bath is considered to be thermalized in WI, one can see that $T\propto \rho_r^{1/4}$ ($\rho_r$ being the radiation energy density during WI), and replacing the radiation energy density by matter energy density, one gets $T\propto \rho_{\rm dm}^{1/4}$. One of the simplest particle physics models of WI is the Warm Little Inflaton model \cite{Bastero-Gil:2016qru} where the light left-handed fermions coupled to the inflaton field yield $\Upsilon_r\propto T$, and the scalar version of this model presented in \cite{Bastero-Gil:2019gao} yields a dissipative coefficient of the form $\Upsilon_r\propto 1/T$. It is in this spirit that the dissipative coefficient of the quintessence field in \cite{Lima:2019yyv} was proposed to have the form shown in Eq.~(\ref{ups-m}). An underlying particle physics model that can yield such a dissipative coefficient for quintessence is yet to be constructed. However, the above form of $\Upsilon_m$ has a nice feature that is, as $\rho_{\rm dm}$ dilutes with expansion, the dissipation of the quintessence becomes more and more prominent. Thus, in this dissipative quintessence model, the dissipation will become effective only at late times. 

With this form of the dissipative coefficient, one can determine how $\rho_{\rm dm}$ evolves in this dissipative DE model. During slow-roll, the EoM of the quintessence field can be approximated from Eq.~(\ref{KG-eqn}) as 
\begin{eqnarray}
3H(1+Q_m)\dot\phi\simeq -V,_\phi,
\end{eqnarray}
inserting which in Eq.~(\ref{dm-dyn}) one gets the evolution equation of $\rho_{\rm dm}$ as 
\begin{eqnarray}
\dot\rho_{\rm dm}+3H\rho_{\rm dm}=\left(\frac{Q_m}{1+Q_m}\right)^2\frac{V,_\phi^2}{\Upsilon_m},
\end{eqnarray}
which, in the weak dissipative regime ($Q_m\ll1$), can be written as 
\begin{eqnarray}
\dot\rho_{\rm dm}+3H\rho_{\rm dm}\simeq Q_m^2\frac{V,_\phi^2}{\Upsilon_m}.
\end{eqnarray}
It is easily seen that in the non-dissipative limit, $\Upsilon_m\rightarrow0$, the above equation yields the standard evolution equation of a non-interactive DM fluid. With the form of the dissipative coefficient given in Eq.~(\ref{ups-m}), one can solve the above equation as 
\begin{eqnarray}
\rho_{\rm dm}&=&\frac{\rho_{\rm dm,0}}{a^3}\left[1+\frac{5M_{\rm diss}^2}{36\rho_{\rm dm,0}^{5/4}}\int \frac{V,_\phi^2}{H^2} e^{\frac{15}{4}\int H dt} dt\right]^{4/5}\nonumber\\
&\equiv&\frac{\rho_{\rm dm,0}}{a^3}\frac{f(\phi/M_p)}{f_0},
\label{rho-dm}
\end{eqnarray}
where $f(\phi)$ has the following form:
\begin{eqnarray}
    f(\phi(a)/M_p)&=& f_0\left[1+\frac{5M_{\rm diss}^2}{36\rho_{\rm dm,0}^{5/4}}\int^a da\, a^{\frac{11}{4}}\frac{V,_\phi^2}{H^3} \right]^{4/5}\nonumber\\
    &\equiv& f_0\left[1+\tilde{C}\,{\mathcal I}(a)\right], \label{fa}
\end{eqnarray}
$\tilde C\equiv 5M_{\rm diss}^2/36\rho_{\rm dm,0}^{5/4}$ being a constant. Again, one can see that in the non-dissipative limit $\Upsilon_m\rightarrow0$ (or equivalently $M_{\rm diss}\rightarrow0$), as $f(\phi/M_p)\rightarrow f_0$ one recovers the standard $1/a^3$ dilution of the DM fluid.\footnote{It is to note that we will not deal with strong dissipative regime in this {\it letter}, and will defer the study of strong dissipative regime for a future work.} Moreover, the normalization factor $f_0$ ensures that as $a\rightarrow 1$ the correct DM abundance today can be recovered.

\subsection{Effective EoS in dissipative DE scenario}

From the above discussion, one can infer that the Friedmann equation for this dissipative DE system can be expressed as 
\beq
3M_p^2H^2=\rho_{\rm dm}+\rho_\phi=\frac{\rho_{\rm dm,0}}{a^3}\frac{f(\phi/M_p)}{f_0}+\rho_\phi,\label{friedmann-theory}
\eeq
where $M_p$ is the reduced Planck mass.
However, while confronting observations, DM and DE fluids are treated as two non-interacting energy components for which the Friedmann equation takes the form:
\begin{eqnarray}
3M_p^2H^2=\rho_{\rm DM}+\rho_{\rm DE},
\end{eqnarray}
where $\rho_{\rm DM}$ scales as $1/a^3$, and $\rho_{\rm DE}$ evolves as 
\begin{eqnarray}
\dot\rho_{\rm DE}+3H(1+w_{\rm eff})\rho_{\rm DE}=0,
\label{DE-evolution}
\end{eqnarray}
where $w_{\rm eff}$ is the EoS that the DESI observations claim to be crossing the phantom-divide. 
Thus, for these two non-interacting fluids, the Friedmann equation can be written as 
\begin{eqnarray}
3M_p^2H^2=\rho_{\rm DM}+\rho_{\rm DE}  = \frac{\rho_{\rm dm,0}}{a^3}+\rho_{\rm DE}.
\label{friedmann-obs}
\end{eqnarray}
Therefore, comparing Eq.~(\ref{friedmann-theory}) and Eq.~(\ref{friedmann-obs}), we see that the effective energy density of DE in this dissipative scenario, as far as the observations are concerned, turns out to be 
\begin{eqnarray}
\rho_{\rm DE}=\frac{\rho_{\rm dm,0}}{a^3}\left\{\frac{f(\phi/M_p)}{f_0}-1\right\}+\rho_\phi. \label{eff-DE}
\end{eqnarray}
Inserting this expression back into Eq.~\ref{DE-evolution}, it is straightforward to show that 
\begin{eqnarray}
1+w_{\rm eff}=\frac{\frac{\rho_{\rm dm,0}}{a^3}\left\{\frac{f(\phi/M_p)}{f_0}-1\right\}+(1+w_\phi)\rho_\phi}{\frac{\rho_{\rm dm,0}}{a^3}\left\{\frac{f(\phi/M_p)}{f_0}-1\right\}+\rho_\phi}.
\end{eqnarray}
One can verify that in the non-dissipative limit when $f(\phi/M_p)\rightarrow f_0$, one gets $w_{\rm eff}\rightarrow w_\phi$, i.e., in the non-dissipative case, the EoS of the quintessence field is what DESI observes and infers as the EoS of DE. Defining 
\begin{eqnarray}
x\equiv -\frac{\rho_{\rm dm,0}}{\rho_\phi a^3}\left\{\frac{f(\phi/M_p)}{f_0}-1\right\},
\end{eqnarray}
the effective EoS of DE in this dissipative DE scenario can be expressed as 
\begin{eqnarray}
w_{\rm eff}=\frac{w_\phi}{1-x}, \label{w-eff}
\end{eqnarray}
which again ensures that the effective EoS of DE differs from the EoS of the dissipative quintessence field for which $x\neq 0$.

\section{Dissipative DE confronting DESI results}
\label{sec:results}

To test the dissipative DE scenario discussed above against the DESI results, we choose a representative inverse-power law potential for the quintessence field as
\begin{eqnarray}
    V(\phi)=V_0\left(\frac{M_p}{\phi}\right)^\alpha.
\end{eqnarray}
In Appendix~\ref{numerical}, we have furnished the methodology to implement this scenario into a numerical code. Here, we will furnish the dynamics of the model, as well as show that the model indeed yields an effective EoS for DE that crosses the phantom-divide as has been observed by DESI. 

\begin{center}
\begin{figure}[!htb]
\subfigure[\, Evolution of the EoS of the quintessence field ]{\includegraphics[width=6.0cm]{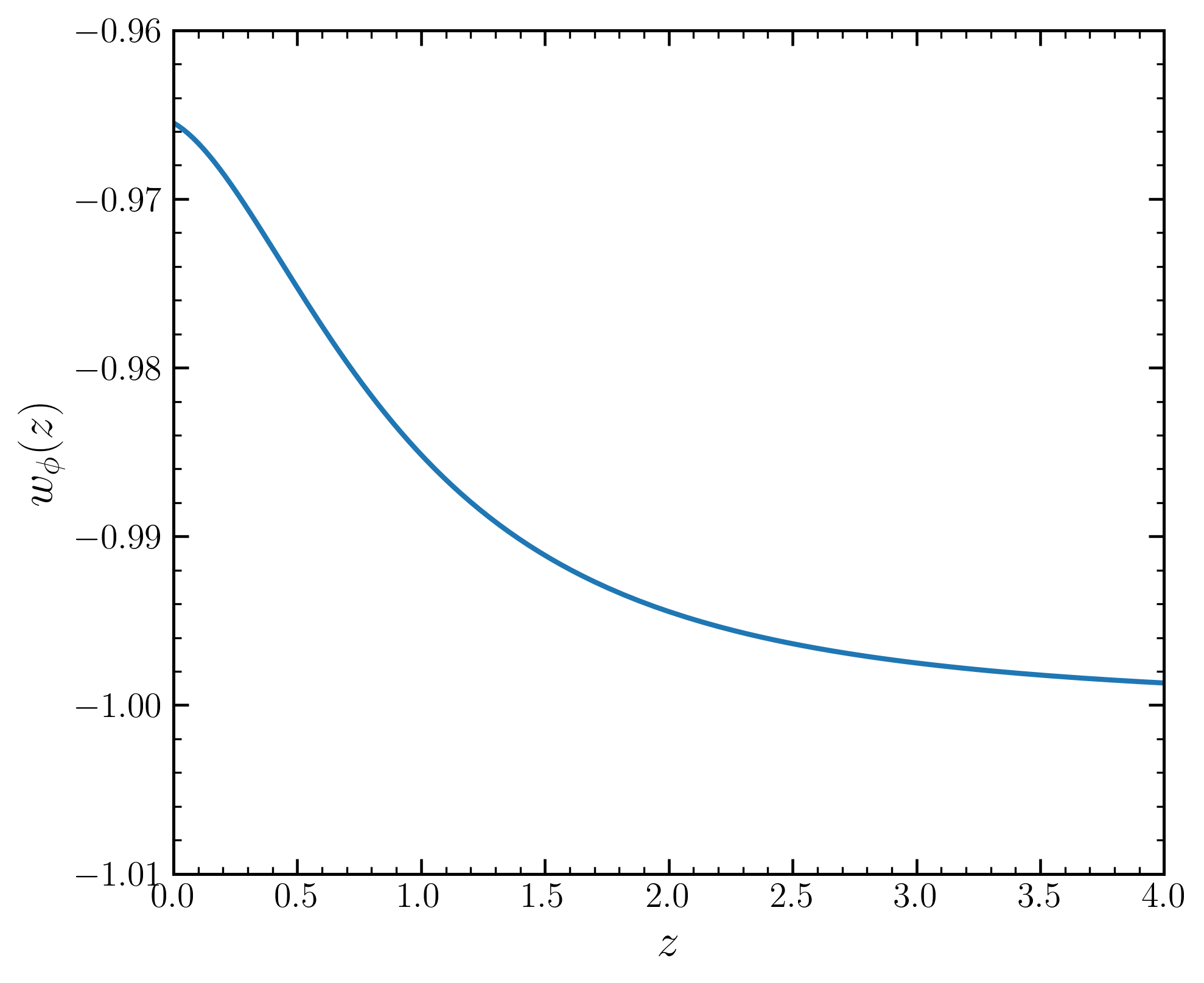}}
\subfigure[\, Evolution of $Q_m$]{\includegraphics[width=5.8cm]{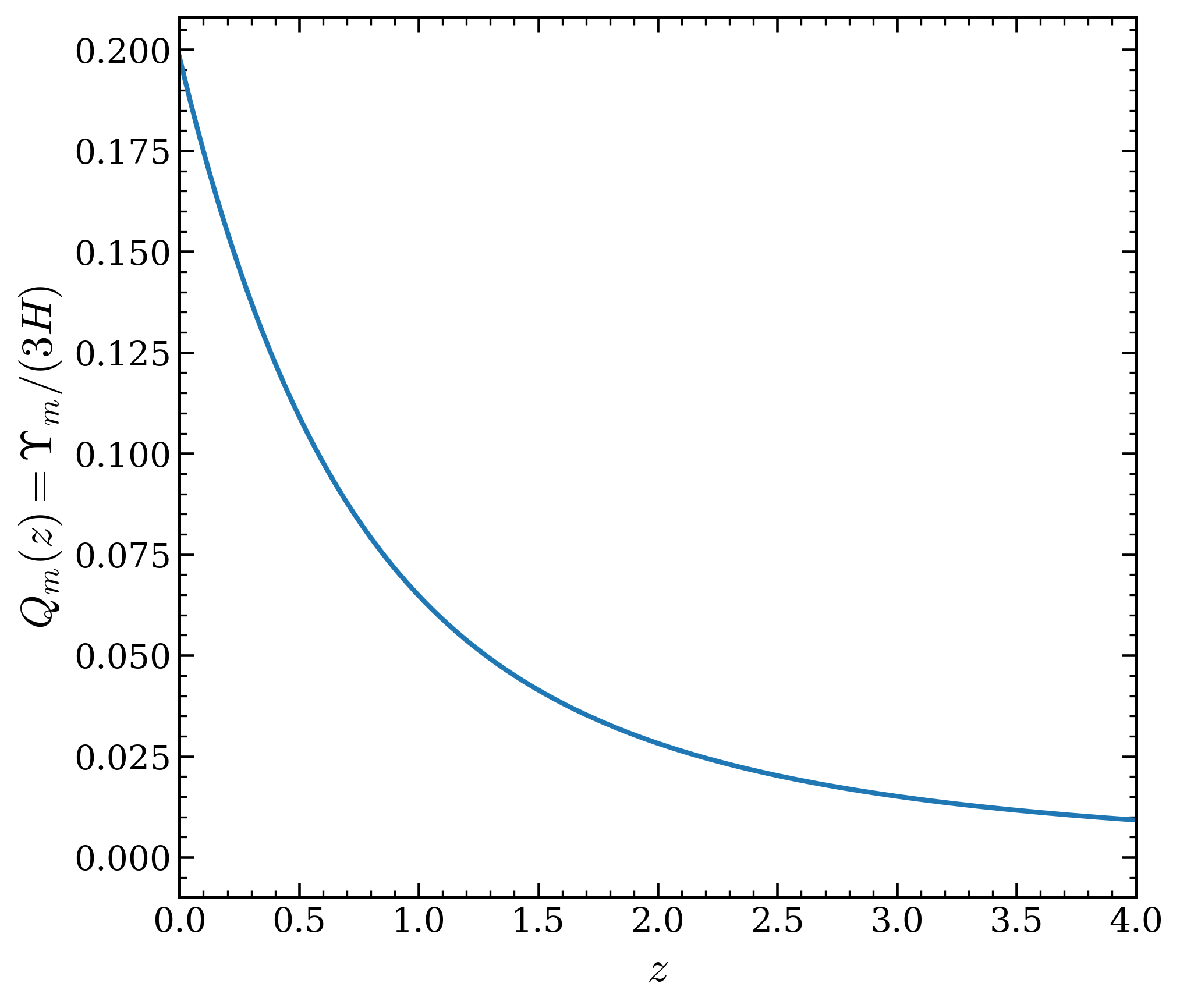}}
\caption{Dynamics of the dissipative quintessential model with inverse-power law potential. We have chosen $V_0=2.1\times10^{-12}$ GeV, $\alpha=1/4$, and $M_{\rm diss}=1.2\times 10^{-27}$ GeV. The figure at the top displays the evolution of the EoS of the quintessence field, and the figure below it shows the evolution of $Q_m$ defined in Eq.~(\ref{Qm}), ensuring that the dynamics are taking place in a weak dissipative regime.}
\label{fig1}
\end{figure}
\end{center}

In Fig.~(\ref{fig1}), we show the evolution of the EoS of the quintessence field, $w_\phi$, on the left-hand side figure and the evolution of $Q_m$ defined in Eq.~(\ref{Qm}) on the right-hand side figure. The evolution of $w_\phi$ shows that this quintessence field never displays a phantom-like behavior as $w_\phi$ never crosses $-1$. The evolution of $Q_m$ shows that $Q_m$ remains smaller than unity throughout the evolution, ensuring that the dissipation is weak throughout, an assumption that we considered while deriving our equations in the previous section. Moreover, we see that $Q_m$ increases at late times, indicating that the dissipation is more effective at late times, leaving the early epochs unaffected. 

\begin{widetext}
\begin{center}
\begin{figure}[!htb]
\subfigure[\, Evolution of the EoS of the effective DE that crosses the phantom-divide ]{\includegraphics[width=8.5cm]{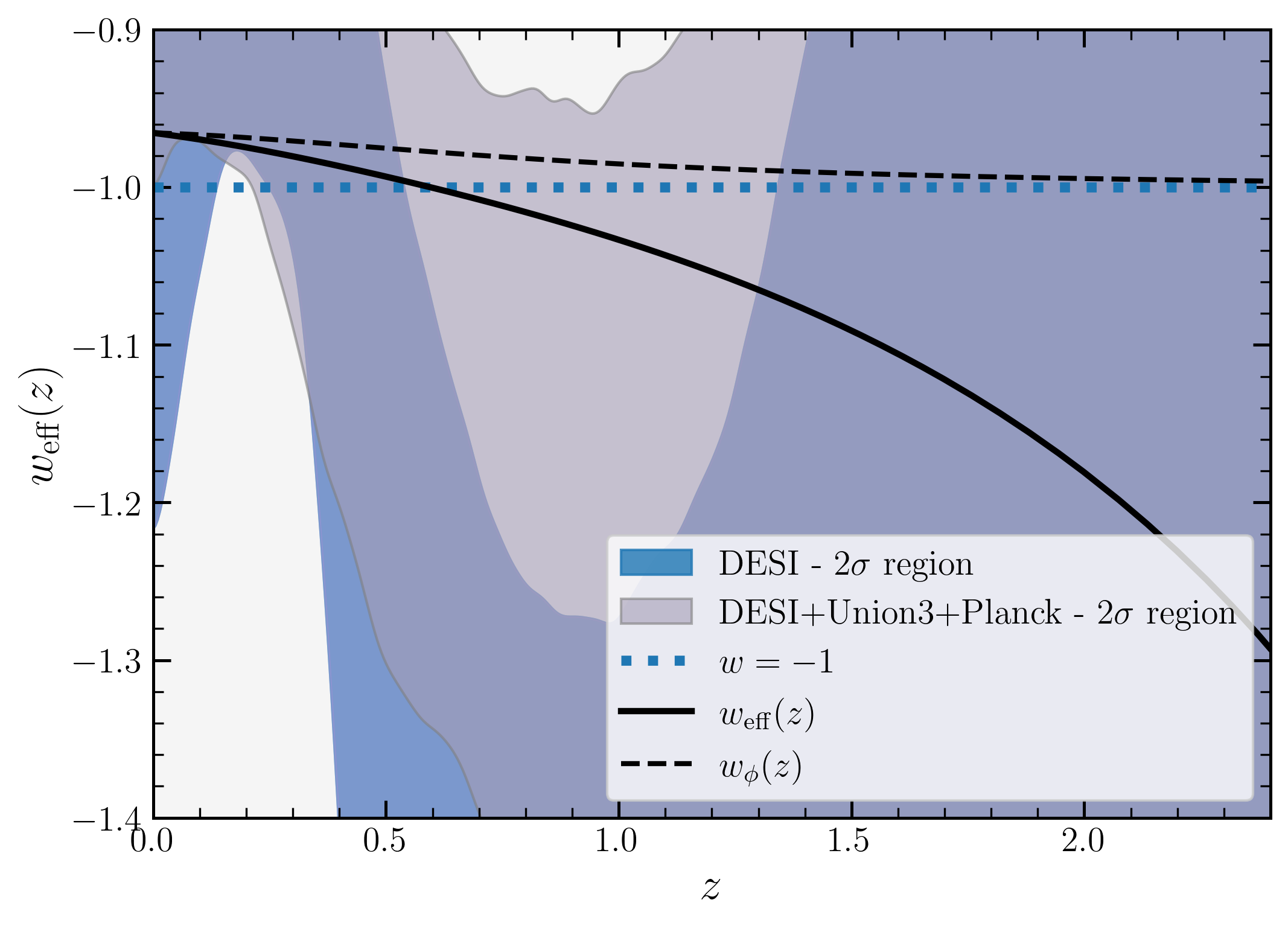}}
\subfigure[\, Evolution of normalized energy density of the effective DE ]{\includegraphics[width=8.5cm]{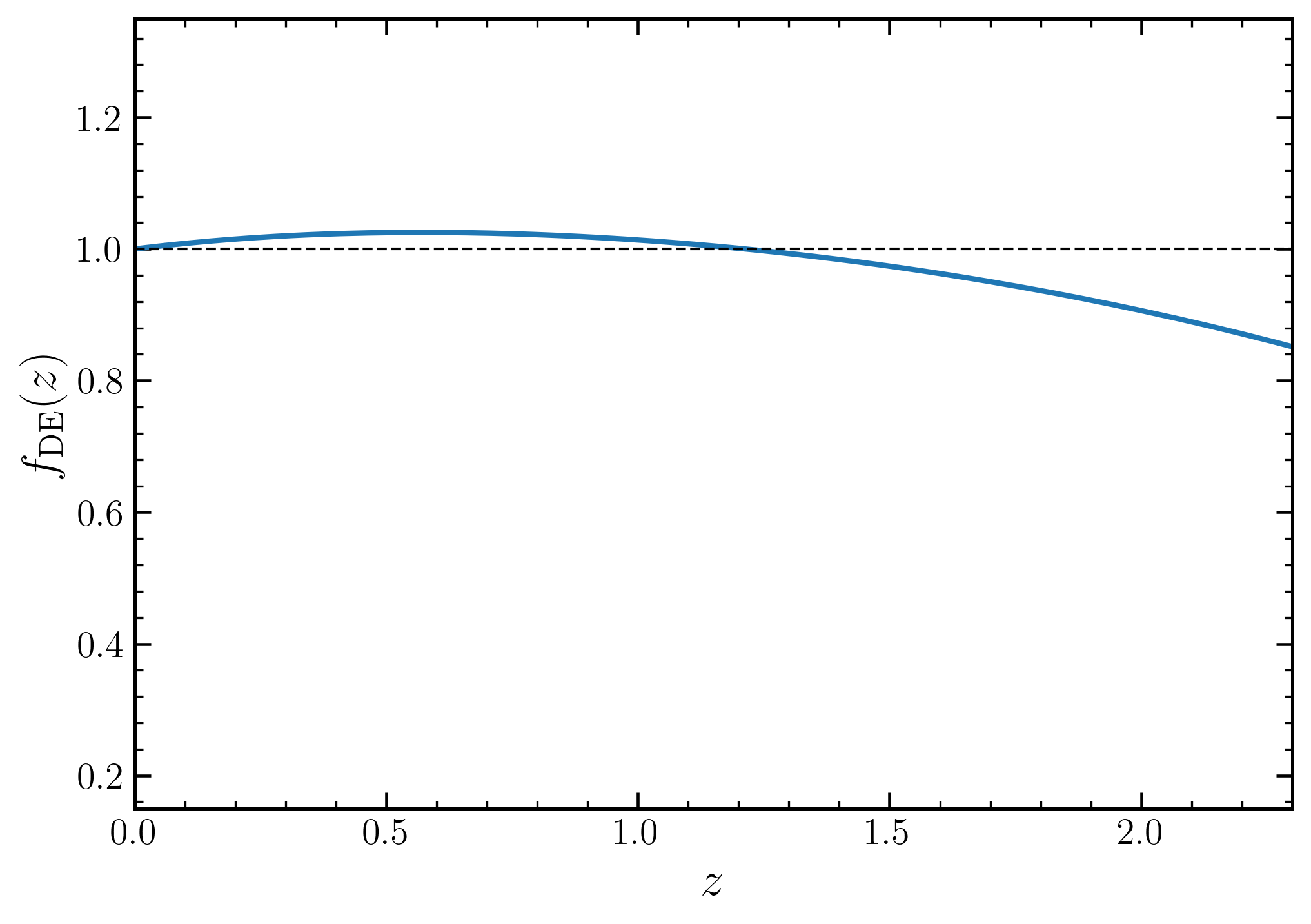}}
\caption{Evolutions of effective EoS ($w_{\rm eff}$) and normalized energy density $(f_{\rm DE})$ of the effective DE fluid in dissipative DE model} 
\label{fig2}
\end{figure}
\end{center}
\end{widetext}

The behavior of the effective DE fluid in this dissipative DE model has been portrayed in Fig.~\ref{fig2}. The left-hand side figure shows that the effective EoS ($w_{\rm eff}$) of the effective De fluid, as has been given in Eq.~\ref{w-eff}, does cross the phantom-divide in the recent past while the EoS of the quintessence field ($w_\phi$) shows no phantom-like behavior. It also shows that the dissipative DE scenario studied here is in accordance with the current DESI results. 

The DESI results also display the evolution of energy density of the (non-interacting) DE fluid normalized to its current value, $f_{\rm DE}(z)$, defined as 
\begin{eqnarray}
    f_{\rm DE}(z)\equiv \frac{\rho_{\rm DE}(z)}{\rho_{\rm DE,0}}.
\end{eqnarray}
For a CC that is non-evolving, the right-hand side will yield 1, and any deviation from 1 would indicate an evolving DE fluid. In the dissipative DE model, the effective DE density is given by Eq.~(\ref{eff-DE}), and the evolution of its normalized form is displayed in the right-hand side of Fig.~\ref{fig2}, indicating that the effective DE fluid in this scenario will appear as an evolving DE fluid for DESI observations. 

\section{Discussion and Conclusion}
\label{sec:summary}

The recent DESI results \cite{DESI:2024uvr, DESI:2024lzq, DESI:2025zgx}, if they stand the test of time, will not only indicate towards an evolving DE component, but also will claim that the DE was phantom-like in the recent past. These results are puzzling mainly because phantom-like dynamics being theoretically pathological is hard to conceptualize. Though standard quintessence dynamics (slow-roll of a real scalar field) can yield an evolving DE component, it doesn't behave like a phantom within GR. However, it was shown a couple of decades ago \cite{Das:2005yj} that an interacting quintessence field, where the quintessence is coupled with the CDM component through interactions, can appear to behave like a phantom field in the past when the system is analyzed as to be non-interacting, as is done while interpreting the DESI results, though the actual dynamics of the quintessence field is not phantom-like ever.  

In this {\it letter}, we showed that a dissipative quintessence field, that dissipates its energy to the concurrent CDM component, can display a similar behavior to an interacting quintessence field and can appear to be phantom-like in the past if the system is interpreted as a non-dissipative one. Dissipation of the inflaton field to a concurrent subdominant radiation bath is a key feature of Warm Inflation \cite{Berera:1995ie}, a variant inflationary scenario. Similar dissipative dynamics for the quintessence field have also been explored in the literature \cite{Lima:2019yyv, Das:2025teu}. The difference between an interacting quintessence field and a dissipative one is that the interactions with CDM in the interaction picture modify the potential of the quintessence field \cite{Das:2005yj}, while the dissipation in the dissipative scenario brings in an extra friction term (apart from the Hubble friction that is already present due to the background expansion) to the quintessence EoM. 

A couple of attractive features of this dissipative quintessence scenario is that, first of all, the dissipation is only effective at late times, given the form of the dissipative coefficient in Eq.~(\ref{ups-m}), and thus is unlikely to affect the CMB and LSS data, unlike the case of the interacting quintessence scenario \cite{Chakraborty:2025syu}, and secondly, a small amount of dissipation is effective enough to make such a scenario compatible with DESI results. However, it is worth keeping in mind that a particle physics model is yet to be developed that can yield such a form of the dissipative coefficient. Moreover, a thorough MCMC analysis is called for to constrain the parameter space for such a dissipative quintessence model to show its robustness. It would further be interesting to explore the dissipative quintessence model in a strong dissipative regime, as has been explored in \cite{Das:2025teu}, to assess its validity given the current DESI results. We leave such a study for future work.

\acknowledgements

Both SDs would like to acknowledge the Workshop on High Energy Physics Phenomenology (WHEPP) XVIII held at IIT Hyderabad in 2025, for providing the platform where this project was discussed for the first time during the cosmology working group activities. Suratna Das would like to thank Swagat S. Mishra for the very useful discussions on DESI results.


\appendix
\section{Implementing the dissipative DE model into a numerical code}
\label{numerical}

In a numerical code, it is easier to cast the equations in terms of the number of $e$-folds ($dN=d\ln a$) rather than in cosmic time $t$. We will further define 
\begin{eqnarray}
    u=\frac{d\phi}{dN}, \label{u}
\end{eqnarray}
and write the quintessence EoM in terms of $u$ rather than in terms of $\phi$. With these redifnitions, the energy density and pressure of the quintessence field $\phi$ can be written as 
\begin{eqnarray}
    \rho_\phi&=&\frac12 H^2u^2+V(\phi), \nonumber\\
    p_\phi&=&\frac12 H^2u^2-V(\phi), \label{rho-p-u}
\end{eqnarray}
and the EoS of the quintessence field is already defined in Eq.~(\ref{w-phi}). The EoM of the quintessence field given in eq.~\eqref{KG-eqn} can then be rewritten as 
\begin{eqnarray}
    \frac{du}{dN}+\left(3+\frac{\Upsilon_m}{H}+\frac{d\ln H}{dN}\right)u+\frac{V,_\phi}{H^2}=0. \label{dudN}
\end{eqnarray}
Considering the contributions of the baryonic matter, $\rho_b$, along with $\rho_{\rm dm}$ and $\rho_\phi$, the Hubble parameter $H$ can be written as 
\begin{eqnarray}
    H^2=\frac{\rho_b+\rho_{\rm dm}+V(\phi)}{3M_p^2-u^2}, \label{Hu}
\end{eqnarray}
and its evolution equation looks like 
\begin{eqnarray}
    \frac{d\ln H}{dN}=-\frac{1}{2M_p^2}\frac{H^2u^2+\rho_b+\rho_{\rm dm}}{H^2}. \label{dHu}
\end{eqnarray}
We further need to determine $f(\phi/M_p)$ as given in Eq.~(\ref{fa}), for which we need to evaluate 
\begin{eqnarray}
    \frac{d{\mathcal I}}{dN}=a^{\frac{14}{5}}\frac{V,_\phi^2}{H^3}. \label{dIdN}
\end{eqnarray}

In a numerical code, one needs to simultaneously solve Eq.~(\ref{u}), Eq.~(\ref{dudN}) and Eq.~(\ref{dIdN}) given the expressions in Eq.~(\ref{rho-p-u}), Eq.~(\ref{Hu}) and Eq.~(\ref{dHu}) along with the forms of $\Upsilon_m$ and $V(\phi)$. The numerical code is initialized at $a_{\rm ini}=10^{-3}$. The initial value of the quintessence field $\phi_{\rm ini}$ is chosen such that its potential energy $V(\phi)$ is approximately of the order of the current DE density. That results in 
\begin{eqnarray}
\phi_{\rm ini}= M_p\left(\frac{V_0}{\rho_{\rm DE,0}}\right)^{1/\alpha}.
\end{eqnarray}
We set the other two initial conditions as $u_{\rm ini}=0$ and ${\mathcal I}_{\rm ini}=0$. 

Moreover, as the normalization $f_0$ in Eq.(\ref{fa}) is not a priori known, the system of equations is solved in two stages for internal consistency. In the first stage, assuming $f_0=1$, one solves the system of equations stated above to find a `raw' solution of $f(a)$, which we denote as $f_{\rm raw}(a)$. Then the normalization factor $f_0$ is determined from $f_0=f_{\rm raw}(a=1)$. In the second stage, the system of equations is again solved with the value of $f_0$ obtained in the first stage while ensuring that $\rho_{\rm dm}(a=1)=\rho_{\rm dm,0}$. We note that in the weak dissipation limit $Q_m\ll1$, the differences between the results obtained in the first and second stages are insignificant. Nevertheless, the second stage ensures full internal consistency of the cosmological evolutions and is therefore retained throughout our analysis. 

\label{Bibliography}
\bibliography{refs}



\end{document}